\documentclass[twocolumn]{aastex62}
\graphicspath{{./}{figures/}}

\usepackage{amsmath}
\usepackage{relsize}
\usepackage{xcolor}

\usepackage{ragged2e}

\usepackage{color}
\definecolor{dodgerblue}{HTML}{1E90FF}
\definecolor{balloonblue}{HTML}{2B60DE}
\usepackage{hyperref} 
\hypersetup{
    colorlinks=true,
    linkcolor=dodgerblue,
    urlcolor=dodgerblue,
    linktoc=all,
    citecolor=balloonblue
           }

\begin{document}
\title{Analysis of a JWST NIRSpec Lab Time Series: Characterizing Systematics, Recovering Exoplanet Transit Spectroscopy, and Constraining a Noise Floor}
\author[0000-0003-4408-0463]{Zafar Rustamkulov}
\email{zafar@jhu.edu}

\affil{Department of Earth and Planetary Science, Johns Hopkins University, 3400 N. Charles Street, Baltimore, MD 21218, USA}

\author[0000-0001-6050-7645]{David K. Sing}
\affil{Department of Earth and Planetary Science, Johns Hopkins University, 3400 N. Charles Street, Baltimore, MD 21218, USA}
\affil{Department of Physics and Astronomy, Johns Hopkins University, 3400 N. Charles Street, Baltimore, MD 21218, USA}

\author[0000-0003-0685-3525]{Rongrong Liu}
\affil{Department of Physics and Astronomy, Johns Hopkins University, 3400 N. Charles Street, Baltimore, MD 21218, USA}

\author[0000-0001-6542-999X]{Ashley Wang}
\affil{Department of Physics and Astronomy, Johns Hopkins University, 3400 N. Charles Street, Baltimore, MD 21218, USA}

\begin{abstract}
The James Webb Space Telescope's NIRSpec instrument will unveil the nature of exoplanet atmospheres across the wealth of planet types, from temperate terrestrial worlds to ultrahot Jupiters. In particular, the 0.6-5.3 $\mu$m PRISM mode is especially well-suited for efficient spectroscopic exoplanet observations spanning a number of important spectral features. We analyze a lab-measured NIRSpec PRISM mode Bright Object Time Series (BOTS) observation from the perspective of a JWST user to understand the instrument performance and detector properties. We create two realistic transiting exoplanet time series observations by performing injection-recovery tests on the lab-measured data to quantify the effects of real instrument jitter, drift, intrapixel sensitivity variations, and 1/$f$ noise on measured transmission spectra. By fitting the time series systematics simultaneously with the injected transit, we can obtain more realistic transit depth uncertainties that take into account noise sources that are currently not modeled by traditional exposure time calculators. We find that sources of systematic noise related to intrapixel sensitivity variations and PSF motions are apparent in the data at the level of a few hundred ppm, but can be effectively detrended using a low-order polynomial with detector position. We recover the injected spectral features of GJ 436 b and TRAPPIST-1 d, and place a 3-$\sigma$ upper limit on the detector noise floor of 14 ppm. We find that the noise floor is consistent with $<$10 ppm at the 1.7-$\sigma$ level, which bodes well for future observations of challenging targets with faint atmospheric signatures.

\end{abstract}


\section{Introduction}

The James Webb Space Telescope (JWST), now in its L2 orbit, will soon shed new light on transiting exoplanet atmospheres. It will probe at high signal-to-noise ratio (SNR) planets spanning the phase space of mass, radius, temperature, and age, yielding our most comprehensive view yet. Recent discoveries by the Transiting Exoplanet Survey Satellite (TESS) \citep{Ricker2015} are beginning to add to an ever-growing catalog of nearby exoplanets that are excellent targets for efficient spectral observations with JWST \citep[e.g.,][]{Winters2019, Jenkins2020, Hobson2021, vanderspek2019}. A variety of spectral modes are available with NIRSpec, each with unique wavelength coverage and spectral resolution. NIRSpec's versatility made it an extremely popular mode for Cycle 1 GO proposers, with roughly two-thirds of awarded exoplanet time series programs making use of one or more of its modes\footnote{https://www.stsci.edu/jwst/science-planning/user-committees/jwst-users-committee}. Several targets on the Guaranteed Time Observations (GTO) and Early Release Science (ERS) lists also widely make use of NIRSpec.

NIRSpec's low-resolution ($\lambda/\Delta\lambda$ $\sim$ 20-100) PRISM mode has a uniquely broad spectral range (0.6 - 5.3 $\mu$m), offering a doubling in throughput relative to other JWST mode combinations, which require a minimum of two separate transits to achieve the same wavelength coverage. This property makes PRISM the ideal instrument to carry out efficient, `one-shot' measurements of a number of carbon, oxygen, and even nitrogen-bearing molecules in exoplanet atmospheres, in all three of the main observational geometries \citep[i.e.][]{Batalha2017, Morley2017b, Lustig-Yaeger2019}. Measurements of these information-rich chemical species will allow for exquisite comparative studies between planets, and will provide important insights into the mechanisms that sculpt planetary atmospheres and populations.

In this work we used a 3-hour Bright Object Time Series (BOTS) dataset measured from the NIRSpec instrument in the lab to look at the data from the perspective of a JWST observer to quantify how the instrument's systematics and noise performance affect the final scientific product. The Pandexo \citep{Batalha2017b} and Pandeia \citep{Pontoppidan2016} packages have served as excellent community tools for estimating exposure times and building synthetic JWST observations. We build off the recent work by \cite{Birkmann2022} and \cite{Jakobsen2022}, who characterize many important engineering-side detector diagnostics, and show how these nuances will affect a real exoplanet observation. Similar characterization work with NIRCam has shown that the detector's noise floor from known systematics sources is 9 ppm, and that its systematics are similar in nature and amplitude to the ones outlined in our work \citep{schlawin2020, schlawin2021}. 

As is often seen in time series data from \textit{Spitzer} IRAC \citep[e.g.,][]{Deming2015, May2020}, \textit{TESS}, and \textit{Kepler}/\textit{K2} \citep{Luger2016, Luger2018}, undersampled PSFs often introduce a non-negligible source of systematic noise due to the PSF's motion across pixels with nonuniform inter- and intrapixel sensitivity variations. Thankfully, with knowledge of the PSF's position over time, generalized pixel-level decorrelation methods (PLD) are able to largely remove these signatures in photometry. With exoplanet time series spectroscopy, analogous jitter decorrelation techniques \cite[e.g.,][]{sing2019} have shown promise in removing systematic trends and greatly improving light curve fits. In our work, we leverage a $\sim$0.3-pixel, $\sim$30-mas drift of the light source on the detector to measure the amplitude of shift-induced systematics in the light curve and the extent to which they can be decorrelated. Further, we fit the observed systematics jointly with the transit light curve at each wavelength channel of our injected transit spectrophotometry, and perform a uniform model intercomparison to predict which systematics vectors are important in the detrending of real JWST time series data. Lastly, we select a wavelength channel that displays weak systematic noise and bin it down through time to place an empirical upper limit on NIRSpec's noise floor.

\section{Overview of the CV3 PRISM Time Series}
We analyze an experimental dataset collected in January of 2016 during the cryo-vacuum testing campaign (CV3) of the JWST Integrated Science Instrument Module (ISIM) at NASA's Goddard Space Flight Center. The single BOTS exposure was taken with NIRSpec held at a temperature of $\sim$40 K (its L2 operating temperature), and consists of 12,000 integrations with three groups each, corresponding to an effective integration time of 0.45 seconds. The exposure used the SUB512 subarray, with size 512$\times$32 pixels, which is wide enough to encapsulate the PSF of the light source and the unilluminated region of the detector. The NRSRAPID readout pattern and S1600A1 (1.6"$\times$1.6") slit were used. The data was processed by the NIRSpec ramp-to-slopes pipeline to fit the up-the-ramp count rate, the statistical variance, and quality flag for each pixel \citep[see][Section 7]{Birkmann2022}. The CV3 BOTS prism data is publicly available through the ESA website\footnote{https://www.cosmos.esa.int/web/jwst-nirspec/test-data}. Full details of the NIRSpec CV3 observation can be found in \cite{Birkmann2022}, which we briefly summarize here. 

\begin{figure}  
    \centering
    \includegraphics[width=\linewidth]{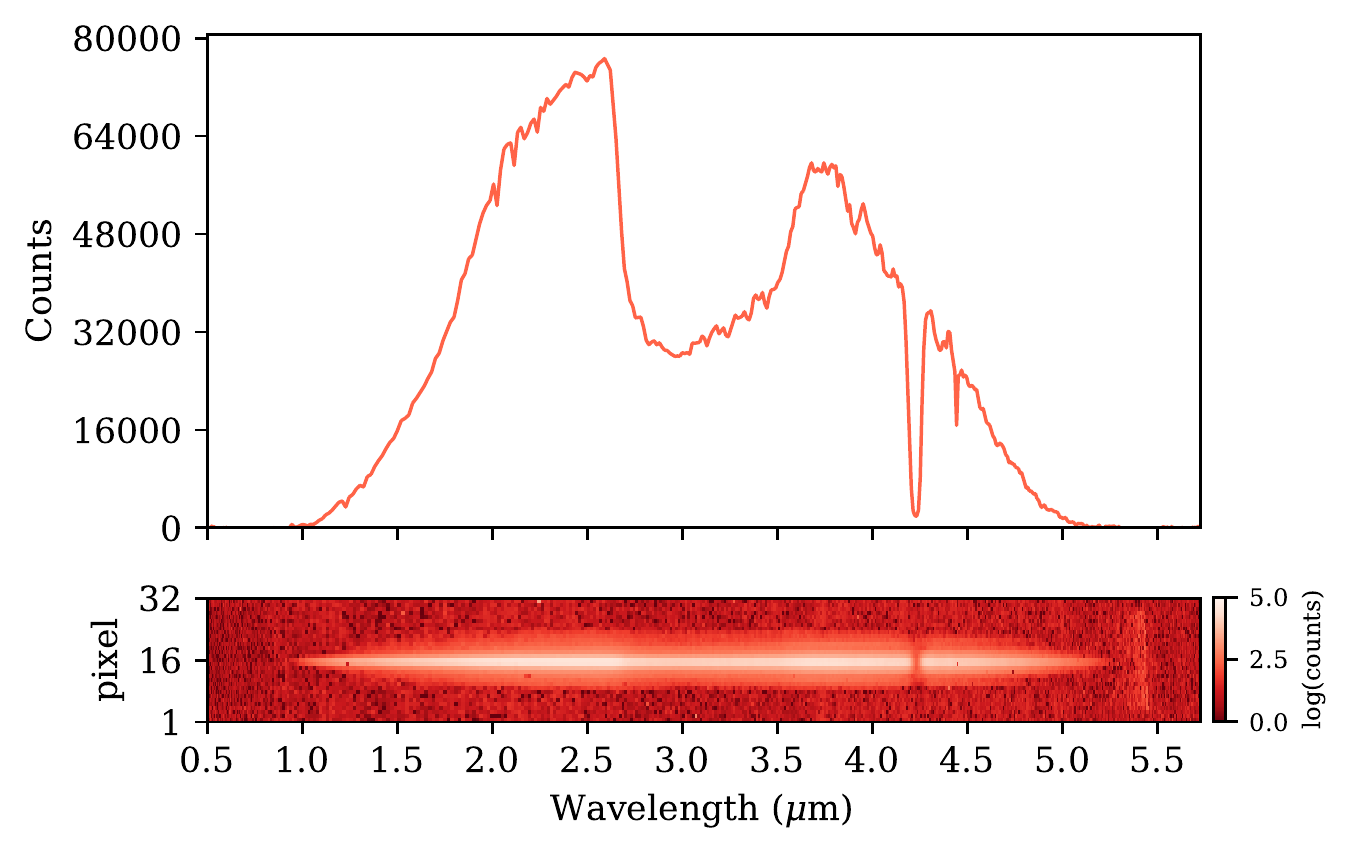}
    \caption{Overview of the lamp spectrum. \textit{Top:} a single 0.45s integration of the lamp spectrum summed along the cross-dispersion axis. Prominent frost absorption features in NIRSpec's optical path sculpt the spectrum's shape. \textit{Bottom:} a single log-stretched raw NIRSpec PRISM frame. The aspect ratio of the detector is vertically exaggerated by a factor of $\sim$2 for clarity.}
    \label{fig:lamp_overview}
\end{figure}

The NIRSpec instrument was illuminated with a tungsten filament source with a similar PSF width as expected from JWST (see Fig. \ref{fig:lamp_overview}). The PSF displays a spatial drift of $\sim$30 mas ($\sim$0.3 px) over the course of the exposure, which $\sim$4$\times$ larger than the 1-$\sigma$ RMS stability requirement for JWST. The light source displays a long term drift in the flux level at the percent-level over several hours, as well as chromatic flickers at the $\sim$0.2$\%$-level, active on the timescale tens of minutes. As is apparent in Fig. \ref{fig:4panel}c, these flickers have a stronger wavelength-dependence at the blue-end of the detector, suggesting that they are related to sudden temperature changes of the filament. The spectrum further shows deep and shallow absorption features presumably due to condensed ices along the optical path. Some of the absorption features change somewhat in depth throughout the exposure, with a notable concave-down profile at 3.55-3.7 $\mu$m that imparts an erroneous excess in the extracted transit depths in this channel. The first hour of the time series shows a strong, chromatic increase in flux due to the lamp warming up, and the chromaticity of the flickers is stronger here. We therefore use the last 2$^{13}$ = 8192 integrations, or $\sim$2.1 hours, of data to mitigate the strong spectral systematics and to better distill the detector's performance. The light source, whose spectrum is far removed from a typical stellar spectrum in shape, has an equivalent J-band magnitude of roughly J$\sim$11, which is approaching PRISM's saturation limit.

\section{Data Extraction and Reduction}\label{Reduction}
\begin{figure*} 
    \centering
    \includegraphics[width=\linewidth]{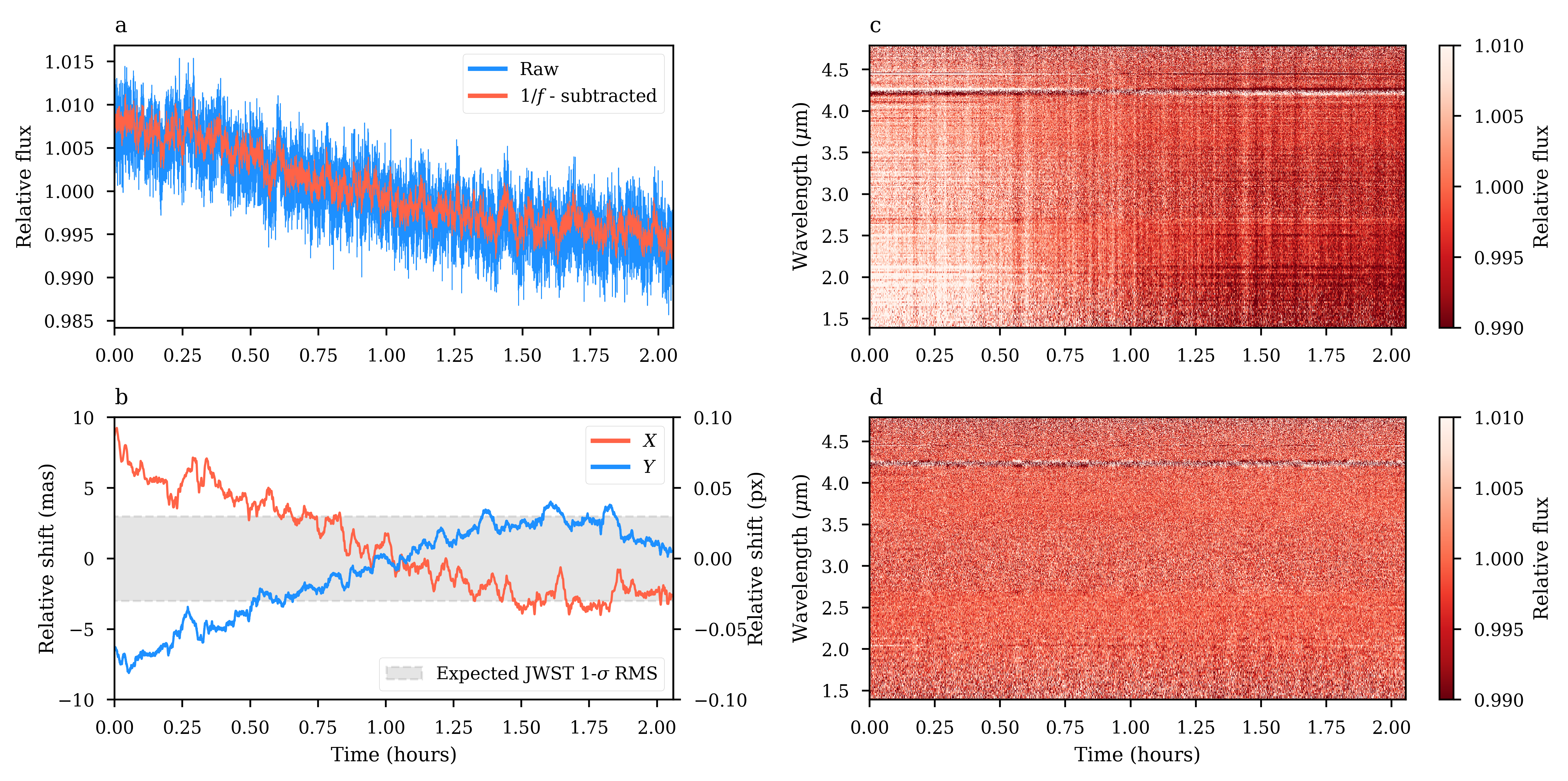}
    \caption{Overview of the NIRSpec PRISM CV3 time series observation. ($\textbf{a}$) The mean-normalized white light curve with and without 1/$f$ correction. (\textbf{b}) The x- and y-shifts of the PSF relative to the middle integration. The grey region shows the 1-$\sigma$ RMS jitter requirement of JWST over 10$^4$ seconds. (\textbf{c}) The raw relative spectrophotometry after shift-stabilization, 1/$f$ noise subtraction, and extraction. Note the strong light source flickers and trends. (\textbf{d}) same as (\textbf{c}), but after common-mode correction that mostly removes the systematics imparted by the unstable light source.}
    \label{fig:4panel}
\end{figure*}

In order to prepare our data for injection-recovery testing, we first extract the spectrophotometry from the time series of the raw frames. The overall approach to reduce and calibrate NIRSpec data has been described by \cite{Rawle2016} and \cite{AlvesdeOliveira2018} and is also summarized in \cite{Birkmann2022} and \cite{Jakobsen2022}.
For each integration, we perform the following steps: we first apply the flat field correction and save the wavelength solution. We next subtract the 1/$f$ noise, as measured by the vertical median of the unilluminated regions of the detector at each pixel column. This process is also referred to as ``destriping." The 1/$f$ noise component, whose contribution can be seen in Fig. \ref{fig:4panel}a, imparts a near-doubling to the white noise level to the spectrophotometry if not correctly removed. Next, we measure the vertical and horizontal pixel shift of the trace in each integration relative to the first by performing a 1-D cross correlation between the vertically- and horizontally-summed counts (Fig. \ref{fig:4panel}b). We then remove these shifts and align spectra by interpolating each frame with a 3rd-order spline and re-gridding the image data onto the shifted pixel array. By aligning the spectra, we remove a common time series systematic effect where the flux of a pixel is observed to change with time as spectral features drift across a pixel.  

Next, we fit a Gaussian profile to the PSF at each pixel column to measure the trace position and width. We smooth the trace profile with a median filter, fit it with a 6th-order polynomial, and perform an sub-pixel extraction of the counts enclosed within 5$\times$FWHM of the trace center. Lastly, we convert from count rate to counts by multiplying by the total integration time of each frame (0.45 s). At this stage, the flat-fielded, 1/$f$ corrected, shift-stabilized, and spatially summed spectrophotometry resembles the true, distilled behavior of the temperamental light source with most of the confounding artifacts removed. The extracted spectrophotometry can be seen in Fig. \ref{fig:4panel}c, where light source flickers and flux variations imparted by x-shifts across sharp spectral features show up strongly in the relative flux at the $\pm$1$\%$-level.

Before injecting transit spectra into the time series, we first apply a common-mode correction to each wavelength pixel in order to suppress the effects of the lamp flicker. This common-mode technique to remove systematic errors is an often-used method to remove time series trends which are the same or similar at each wavelength \citep[e.g.,][]{Berta2012, Deming2013, Sing2013}.
While simply dividing the time series of each pixel by the white light curve can remove most of the observed lamp flickers and long-term flux drifts, the amplitude of these variations are found to also have a color dependence, which we remove by applying the common-mode correction on a pixel-by-pixel basis. This is done by performing a linear regression fit of the white light curve (see Fig. \ref{fig:4panel}a) to the spectral light curves at each wavelength pixel, and dividing it out. This correction removes the majority of the light source flicker while preserving the detector's noise properties, intrapixel systematics, and most of the light source's chromaticity. The self-correction effect imposed by division of the white light curve can only be seen in the post-corrected white light curve, which by definition is a flat line at 1. The common-mode-detrended spectrophotometry is shown in Fig. \ref{fig:4panel}d, where modest residual chromatic light source systematics are seen. At this stage, the spectrophotometry is largely free from confounding trends and flickers from the light source. 

\section{Transit Spectrum Injection}
Injection-recovery tests are often repeated with several draws from a large dataset or noise model, but our study relies on a single real observation, and therefore a single noise instance. We perform our tests on two different planet cases, as two separate manifestations of the same noise instance, to show the consistency of our injection-recovery process. We inject the transit spectrophotometry of two planets: the warm Neptune GJ 436 b (T$_{eq} \sim$700 K, J = 6.9) and the temperate terrestrial planet TRAPPIST-1 d (T$_{eq} \sim$300 K, J = 11.35). We choose these planets due to their short $\leq$1 hr transit duration and their broad interest to the exoplanet atmosphere community.

GJ 436 b is slated for transmission spectroscopy with NIRSpec G395H (Stevenson and Lustig-Yeager et al., GO 1981), and TRAPPIST-1 d will be observed in transit with the NIRSpec PRISM (Lafreniere et al., GTO 1201). With TRAPPIST-1's J-band magnitude of 11.4, the count rates from the CV3 light source are similar in our dataset to what will be seen on-sky with JWST, making it a compelling comparison. GJ 436A, however, is 4 magnitudes brighter (prohibitively bright for PRISM) and the expected precision of its measured transmission spectrum using other modes is significantly higher than what we present. For GJ 436 b, we use a transmission spectrum from the Goyal Grid of generic models \citep{Goyal2020} with parameters most closely matching those of GJ 436 b. The TRAPPIST-1 d spectrum is taken from the work of \cite{Lincowski2018} and  \cite{Lustig-Yaeger2019} where a CO$_2$-dominated 10-bar atmosphere is assumed, with and without high-altitude H$_2$SO$_4$ clouds. 

For each planet, we bin our model transmission spectra to the wavelength grid of each PRISM pixel, and compute 4-parameter limb darkening coefficients between the pixel edges using values from the 3D STAGGER-grid \citep{Magic2015}. These LD coefficients are weighted by NIRSpec's response function and the respective stellar spectrum. Given the narrow wavelength bins used here, we find that the errors in the LD coefficients introduced by using a stellar spectral weighting rather than the lamp's spectrum are very small. We then compute the transit light curve time series at each wavelength pixel using \texttt{batman} \citep{Kreidberg2015}.

Finally, we inject the spectral transit model while simultaneously dividing out the common-mode correction into the raw data, and reinsert the original time series' 1/$f$ noise and jitter. At this stage, our injected files mimic a JWST observation as closely as possible, with the jitter exaggerated by a factor of 4.

\section{Transit Spectrum Recovery} \label{Recovery}
We reduce our transit-injected data files following the procedure described in the first paragraph of Section \ref{Reduction}. We bin the transit spectrophotometry into variable-width wavelength channels with roughly equal counts. We use 51 and 19 wavelength channels for GJ 436 and TRAPPIST-1, respectively. 

We model the systematic errors with a parameterized deterministic model, where the photometric trends are found to correlate with a number of external parameters (or optical state parameters, \textbf{x}). These parameters describe changes in the instrument or other external factors as a function of time during the observations, and are fit with coefficients, $p_n$, for each optical state parameter to detrend the light curves.
We then fit each spectral light curve jointly with a transit model and systematics model $S(\mathbf{x})$ of the form:
\begin{equation}
    S(\mathbf{x}) =  p_1X + p_2Y + p_3X^2 + p_4Y^2 + p_5XY +  p_6\Delta_{\mathrm{CM}}
\end{equation}
where $X$ and $Y$ are the mean-subtracted, standard deviation-normalized x- and y-shifts, respectively, and $\Delta_{\mathrm{CM}}$ is the pre-injection normalized common-mode light curve, normalized in the same way. The full light curve flux model, $f(t)$, is then assembled with the equation

\begin{equation}
f(t)= f_0 \times T(t,\boldsymbol{\theta})\times  S(\mathbf{x})
\end{equation}
where $f_0$ is the baseline flux and $T(t, \theta)$ is the theoretical transit model which depends on the transit parameters $\theta$. This model captures the behavior of jitter-induced intrapixel variations, as well as leftover light source variability. The detrending vectors are normalized so that their fit coefficients may be directly compared. We perform joint fits of the systematics model and the planet's transit depth (while keeping all other transit parameters fixed) using the Python-based Levenburg-Marquardt least squares algorithm \texttt{lmfit} \citep{Newville2014}. Using Python's \texttt{multiprocessing} package to parallelize the fitting routine, we fit all 51 GJ 436 b channels in 30 seconds using 16 cores. We initially assume photon noise uncertainties for the light curve, re-fit with the residual standard deviation as the uncertainty, and then inflate the final \texttt{lmfit} parameter uncertainties on the transit depths by a factor $\beta(\lambda)$. $\beta(\lambda)$ is computed from the fit-residuals of each spectral light curve taking into account both white and red noise sources. Its median value is 1.35, and it peaks where there are residual systematics due to chromatic lamp flickers and changing cryogenic ice features. $\beta(\lambda)$ is near 1 in wavelength regions where these two effects are small. 
If $\beta(\lambda)$ were not multiplied into our transit depth uncertainties, the residual chromatic flicker effects would not be taken into account, giving unrealistic errorbars. Further, we avoid removing these residuals using stochastic models such as Gaussian Processes (GPs) since they could remove systematics due to detector's noise properties which we aim to characterize.

\begin{table*}[h!]
		\centering
	    \caption{Systematics Fitting Results Summary}
		
		\begin{tabular}{cccccccccccc}
			\hline    
			\hline 
			Scenario & Description & $N_\textrm{params}$ & $\sigma_{\mathrm{TS \ resid}}$ (ppm) & $\chi^2_{\nu}$ & $\Delta$BIC & $\Delta$AIC \\
			\hline
			1& Fiducial & 8 & 148.70 & 1.32  & -- & --\\
			2& No $X^2$ & 7 & 148.74 & 1.25  & -6.15 & -2.48 \\
			3& No $XY$ & 7 & 149.30 & 1.31  & -5.01 & -1.35 \\
			4& No $Y^2$ & 7 & 151.21 & 1.28  & -4.87 & -1.20 \\
			 5& No $X^2, Y^2$ & 6 & 152.38 & 1.28 & -7.37 & -0.08 \\
			 6& No $Y$ & 7 & 171.28 & 2.32  & 39.09 & 42.75 \\
			 7& No $X$ & 7 & 184.86 & 2.25  & 36.15 & 39.85 \\
			 8& Fiducial with  $X^3, Y^3$ & 10 & 243.70 & 3.36 & 88.69 & 81.27 \\
			 9& No $X^2, Y^2, XY$ & 5 & 267.5 & 4.33  & 130.17 & 141.02 \\
			 10& No $X^2, Y^2, XY, X$ & 4 & 285.93 & 4.84  & 154.76 & 169.10 \\
			 11& No $X^2, Y^2, XY, Y$ & 4 & 290.39 & 6.05  & 211.69 & 226.03 \\
			 12& No $X^2, Y^2, XY, X, Y$ & 3 & 293.37 & 5.66  & 195.09 & 212.79 \\
			 \hline
			 13& Fiducial, no flat field & 8 & 390.98 & 7.12 & 248.97 & 248.97 \\
			 14& Fiducial, no 1/$f$ subtraction & 8 & N/A & 555.69 & 23837.78 & 223837.78 \\

			\hline 
		\end{tabular}
		\justify{Results from a suite of tests for goodness-of-fit and systematics model quality sorted in decreasing order of residual scatter. Each row represents an independent transmission spectrum extraction from our injected transit spectrophotometry of GJ 436 b, fitted with some linear combination of 1st, 2nd, and 3rd order $X$ and $Y$ jitter vectors, and the cross-term $XY$. $\sigma_{\mathrm{resid}}$ is the standard deviation of the residual between the injected and recovered transmission spectrum, and $\chi^2_{\nu}$ is the reduced $\chi^2$ statistic of the fit. The scatter in Scenario 14 is not reported since the noise level was too overwhelming for the fit to prefer a non-zero transit depth.}
		\label{tab:Intercomparison} 
	\end{table*}

We also use the nested sampling package \texttt{dynesty} to fit the light curves \citep{Speagle2020}, and find equivalent results in both the fitted depth and the uncertainties. Moreover, we find that there are no complicated degeneracies between the transit depth and the optical state vectors, indicating that the posterior is well-described by a multivariate Gaussian distribution. This means that a Levenburg-Marquardt least squares minimizer is sufficient to quickly estimate the posterior distribution with accurate uncertainties.

This fit to the transmission spectra of both planets, which we term the ``fiducial" fit hereafter, match the injected spectra well, with a reduced chi-squared $\chi^2_\nu$ of 1.32 for GJ 436b, and 1.15 for TRAPPIST-1 d (see Fig. \ref{Recovery}). In the latter case, the extracted spectrum qualitatively shows increases in transit depth at the three main absorption features. The $\chi^2_\nu$ for GJ 436 b's fiducial fit without taking into account errorbar inflation is 3.13. The residuals are largely Gaussian, with the notable exception of the peaky 3.6$\mu$m feature, which is imparted by a downward concavity in the ice deposition profile there. This same feature can be seen in Fig. A.2, panel f, of \cite{Birkmann2022}. The average amplitude of the $X$ and $Y$ shift vectors are found to be 455 and 445 ppm, respectively, across both spectra. This is in excellent agreement with the jitter systematic noise modeled by \cite{Birkmann2022}. Rescaling these amplitudes linearly by a factor proportional to the expected jitter of JWST, we estimate the on-sky jitter-induced photometric signatures to be at the $\sim$100 ppm level.

\begin{figure*} \label{spectrum_recovery}
    \centering
    \includegraphics[width=.65\linewidth]{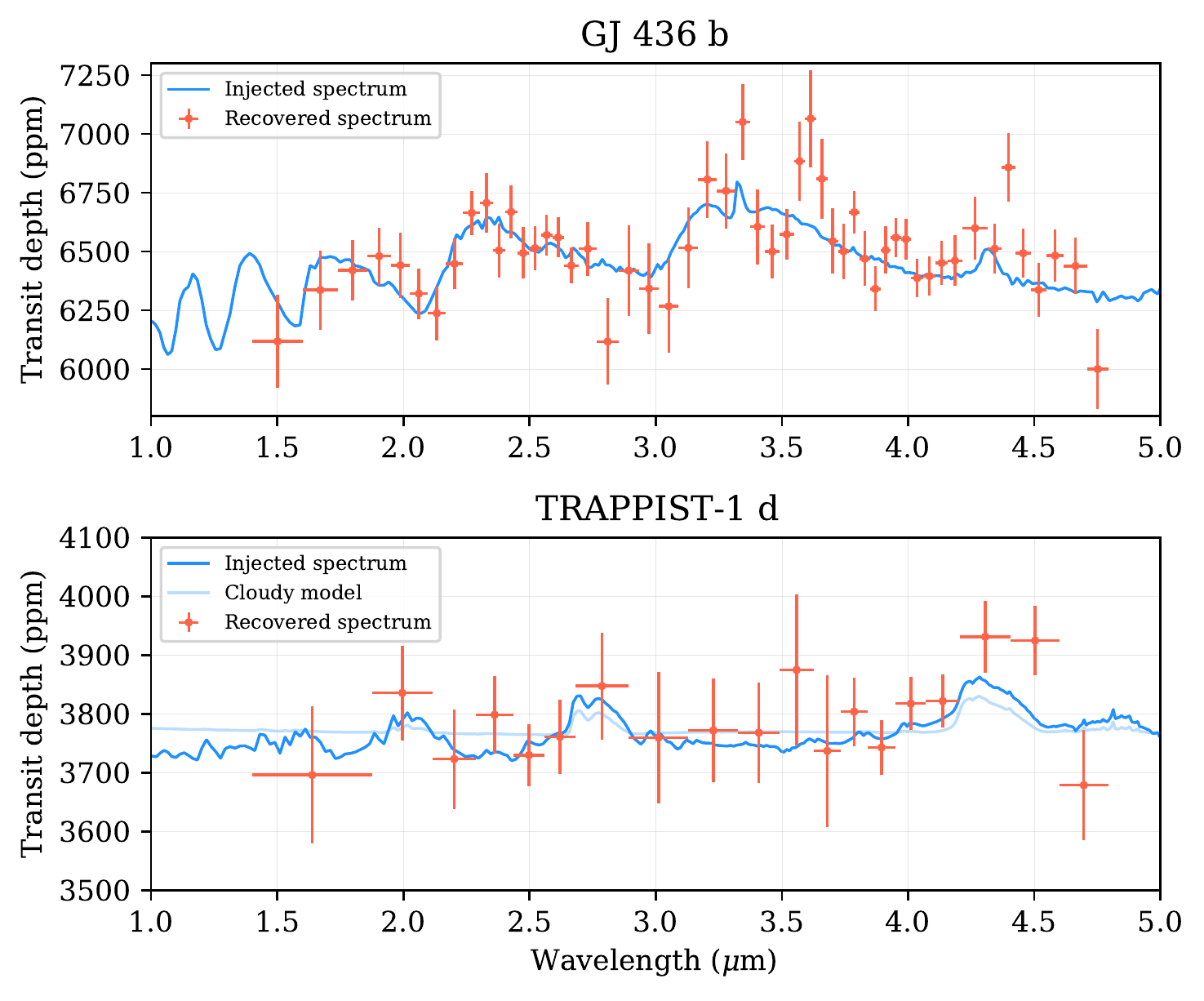}
    \caption{The recovered transmission spectra of GJ 436 b (\textit{top}), and TRAPPIST-1 d (\textit{bottom}) recovered from the injected data using our novel pipeline. We include a cloudy TRAPPIST-1 d atmosphere model for an illustrative comparison.}
    \label{fig:my_label}
\end{figure*}

\section{Systematics Model Intercomparison}
We explore the relevance of each term in the systematics model, as well as the flat field and 1/$f$-subtraction, by performing a model intercomparison on the GJ 436 b data. We re-fit the transit spectrophotometry using the methods described in Section \ref{Recovery}, but with key elements added or removed to see how the goodness-of-fit of the final transmission spectrum changes. We recommend this technique as an informative method for systematics model selection, and to explore the response of the final spectrum to different procedures in the data reduction. At each scenario, we compute the standard deviation of the fit residual, the reduced Chi-squared statistic $\chi^2_\nu$, and the change in Bayesian and Akaike Inference Criteria ($\Delta$BIC and $\Delta$AIC, respectively) relative to the fiducial model. We note that in each instance, we keep the uncertainties on the transit depths fixed to those of the fiducial fit in order to avoid erroneously inflating the errorbars due to poor model selection, and to preserve the uniformity of the comparison. The results of this intercomparison are in Table \ref{tab:Intercomparison}.

Decorrelation against the linear jitter vectors $X$, $Y$, and $XY$ all substantially improve goodness-of-fit over a model without them. The second-order terms $X^2$ and $Y^2$ are marginally, if at all, important, as evidenced by the negative changes to their Bayesian and Akaike Information Criteria relative to the fiducial model ($\Delta$BIC and $\Delta$BIC, respectively). This indicates that the intrapixel sensitivity profile of the typical NIRSpec pixel is well-approximated by a low-order polynomial. Adding the third-order terms $X^3$ and $Y^3$ to the systematics model results in a much poorer goodness-of-fit, likely due to the even nature of the detector's intrapixel sensitivity function.

\begin{figure*} 
    \centering
    \includegraphics[width=0.95\linewidth]{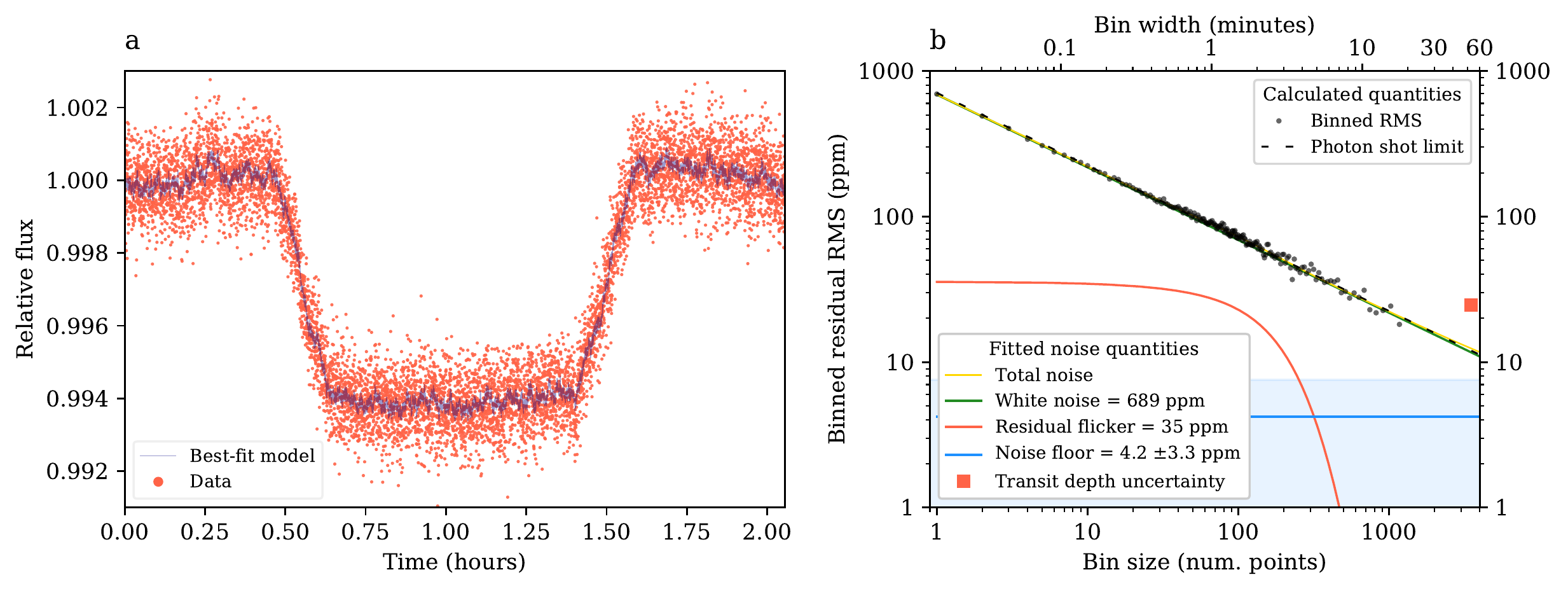}
    \caption{\textbf{(a)} The injected transit light curve photometry of GJ 436 b in the relatively systematics-free 3.77-4.17 $\mu$m channel (tomato points), and the best-fitting joint model (grey line). \textbf{b} Residual scatter as a function of bin size for the 3.77-4.17 $\mu$m channel (black points) with the best-fitting noise components (solid lines) and the photon shot noise limit (dashed line).}
    \label{fig:noisefloor}
\end{figure*}

Flat fielding has a major impact on the quality of the resulting fit, nearly halving the resulting residual scatter when the flat field is applied. From this large effect, we determine that significant efforts should be made to obtain several, accurate flat fields for the subarrays used in JWST BOTS data. Sub-pixel flat fields could help further reduce the impact of jitter on the photometry. Perhaps most important, though, is the 1/$f$ correction--this added noise component overwhelms the transit depths, resulting in non-detections of the transit feature in all spectroscopic channels when not removed.

A number of systematic sources that are not explored by our work, or potentially fully captured by this dataset, are likely to be present in on-sky JWST data. \cite{Birkmann2022} discusses many potential sources of systematic sources in more depth, including
charge trapping-induced persistence in the detector, which results in exponential ``ramps" in measured flux as is regularly seen in HST/WFC3 data (e.g \citealt{Deming2013,Wakeford2013,Kreidberg2014,Zhou2017}). This effect, which cannot be measured in the CV3 data due to the instability of the lamp, is expected to be greatly reduced for the relatively new JWST detectors \citep{rauscher2014, Birkmann2022}. High-gain antenna moves, increased cosmic rays, and thermal drift of the telescope may also introduce systematics and are not captured in the CV3 data. \cite{Birkmann2022} estimates that slit losses and intrapixel sensitivity variations will both be on the order of several hundred ppm with bright object time series data, pending JWST's real on-sky PSF profile and pointing stability. The CV3 data displays these trends at the expected levels given the $X$ and $Y$ position dependence of the spectrophotometry, which are well-decorrelated by our jitter detrending technique.

\section{Searching for the Noise Floor}
To search for the detector's noise floor, we first select a wide wavelength region that is largely free of systematics and light source spectral features. A bin sufficiently narrow to minimize the enclosed chromatic systematics, and wide enough to encapsulate maximal counts, is ideal. The relatively clean 3.77-4.17 $\mu$m region is chosen for these reasons. We then fit the light curve of GJ 436 b at this channel as described previously, and study its residual by binning it through time (see Fig. \ref{fig:noisefloor}). The residual RMS bins down closely with $1/\sqrt{N}$ (where N is the number of points per bin). We fit this RMS curve with a quadrature-sum of the white noise component, a noise floor, and a logistic decay law to capture the leftover chromatic flickers. We find that the fitted white noise at this bin closely matches the expected photon shot limit, and that the residual flicker noise contributes 35 ppm at the $\sim$2-minute timescale, which is consistent with the flicker timescale seen in the white light curve. The RMS curve bins to $\sim$15 ppm. Most importantly, we recover a 3-$\sigma$ upper bound of 14 ppm for the noise floor, with the fit preferring a $<$10 ppm floor at 1.7-$\sigma$. Given the joint probability distribution with our systematics model, the transit depth uncertainty at this channel is 1.35$\times$ larger than what is expected due to photon noise alone, or 74$\%$ of the photon noise limit. We expect on-sky JWST, with its expected weaker systematics, to reach closer to the photon limit, making our value somewhat pessimistic.

\section{Summary and Conclusion}
In this work we analyzed NIRSpec lab time series data, injecting and recovering transmission spectra including real detector noise, making it the closest analog available before science data is collected. The dataset, complete with the detector's actual noise performance and intrapixel sensitivity variations, informed our optimistic expectations as future JWST users. We demonstrate that subtraction of the 1/$f$ noise is a crucial step in the data reduction, and caution that this step may be complicated in the narrower SUB512S subarray which is only 16 pixels tall. Further, we show that neglecting the flat field increases the scatter of the final spectrum by a factor of 2.6 (Table \ref{tab:Intercomparison}). Given JWST's 7 mas (0.07 pixel) on-sky stability requirement we expect the amplitude of jitter-induced systematics to be at the $\sim$100 ppm-level. We demonstrate that such jitter systematics can be readily detrended using a low-order polynomial with detector position, binning the data down close to the photon limit. While we do not confidently detect a noise floor, we show that it is $<$14 ppm, and likely below 10 ppm. Using our fast reduction and analysis pipeline, we successfully recover our injected transmission spectra, and show that Levenburg-Marquardt least-squares fitters are sufficient to quickly estimate the posterior distribution. We also find that, when simulating observations, transit depth uncertainties should be scaled conservatively by a factor of 1.35 relative to the photon noise limit expectation. All of these facts considered, we believe NIRSpec PRISM is well-suited to characterize challenging exoplanet atmosphere targets with $\sim$10 ppm spectral amplitudes. We expect that the precision of such spectra will be limited by the photon limit and confounding astrophysical signatures such as stellar activity rather than instrumental systematics.

\begin{acknowledgments}
We thank Giovanna Giardino, Stephan Birkmann, and Pierre Ferruit for their useful discussions and the NIRSpec team for their efforts in collecting and sharing the CV3 dataset. 
\end{acknowledgments}

\vspace{5mm}
\facilities{JWST(NIRSpec)}

\software{\texttt{Matplotlib} \citep{hunter:2007}, \texttt{Numpy} \citep{vanderwalt:2011}, \texttt{Scipy} \citep{virtanen:2020}, \texttt{astropy} \citep{astropy:2013}, \texttt{dynesty} \citep{Speagle2020}, \texttt{lmfit} \citep{Newville2014}}

\bibliography{Exoplanets_Z_paper}


\end{document}